\documentclass[10pt,a4paper,twocolumn,english,prb,aps,showpacs,floatfix,groupedaddress,superscriptaddress]{revtex4-1}
\usepackage{graphicx}
\usepackage{epsfig}
\usepackage[english]{babel}
\usepackage{amsmath}
\usepackage{amssymb}
\usepackage{amsfonts}
\usepackage{longtable}
\usepackage{dcolumn}
\usepackage{bm}
\usepackage{bbm}
\usepackage{nicefrac}
\usepackage{color,array}
\usepackage{colortbl}
\usepackage{dsfont}
\usepackage{mathtools}
\usepackage{subfigure}  
\usepackage{soul}

\usepackage{xcolor,cancel}

\def\ket#1{\mathinner{|{#1}\rangle}}           

\begin{document}

\newcommand{\be}{\begin{equation}}
\newcommand{\ee}{\end{equation}}
\newcommand{\bearr}{\begin{eqnarray}}
\newcommand{\eearr}{\end{eqnarray}}
\newcommand{\bseq}{\begin{subequations}}
\newcommand{\eseq}{\end{subequations}}
\newcommand{\nn}{\nonumber}
\newcommand{\reqn}{\eqref}

\title{Orientational bond and N\'eel order 
in the two-dimensional ionic Hubbard model}

\author{Mohsen Hafez-Torbati}
\email{mohsen.hafez@tu-dortmund.de}
\affiliation{Lehrstuhl f\"ur Theoretische Physik I, 
Technische Universit\"at Dortmund,
Otto-Hahn-Stra\ss e 4, 44221 Dortmund, Germany}

\author{G\"otz S. Uhrig}
\email{goetz.uhrig@tu-dortmund.de}
\affiliation{Lehrstuhl f\"ur Theoretische Physik I, 
Technische Universit\"at Dortmund,
Otto-Hahn-Stra\ss e 4, 44221 Dortmund, Germany}

\date{\rm\today}

\begin{abstract}
Unconventional phases often occur where two competing mechanisms compensate. 
An excellent example is
the ionic Hubbard model where the alternating local potential 
$\delta$, favoring a band insulator (BI), competes with the local repulsion $U$,
favoring a Mott insulator (MI).
By continuous unitary transformations we derive effective models in which
we study the softening of various excitons. The softening signals
 the instability towards new phases that we describe on the mean-field level.
On increasing $U$ from the BI in two dimensions, we 
find a bond-ordered phase breaking orientational symmetry
due to a d-wave component. Then, antiferromagnetic order appears coexisting
with the d-wave bond order. Finally, the d-wave order vanishes and 
a N\'eel-type MI persists.
\end{abstract}



\maketitle

Searching for unconventional states of matter and 
non-trivial elementary excitations (quasiparticles (QPs)) 
is one of the crucial objectives of the research
in strongly correlated lattice models.
Outstanding examples range from the quasi long-range ordered Mott insulator (MI)
in one dimension (1D) where the neutral spin-1/2 particle ``spinon'' represents the elementray excitation to quantum spin ice in three dimensions 
(3D) where magnetic monopoles represent the QPs. 
In order to find unexpected phases, it is a good idea to focus
on parameter regions where two antagonists compensate
because then subtle subleading mechanisms can take over.

The present article addresses the IHM in two dimensions (2D) on the
square lattice
in order to understand which phases possibly arise between the band
insulator (BI) and the MI.
Its Hamiltonian reads
\begin{eqnarray}
H=\frac{\delta}{2}\sum_{\bm{ r},\sigma} (-1)^{i+j} 
n^{\phantom{\dagger}}_{\bm{ r},\sigma} 
&+&U\sum_{\bm{ r}} ( n^{\phantom{\dagger}}_{\bm{ r},\uparrow}-\frac{1}{2} ) 
(n^{\phantom{\dagger}}_{\bm{ r},\downarrow}-\frac{1}{2} ) \nn 
\\
&+&t\sum_{\langle\bm{ r}\bm{ r'}\rangle,\sigma} 
(c^\dagger_{\bm{ r},\sigma} c^{\phantom{\dagger}}_{\bm{ r}',\sigma} + 
{\rm h.c.}),
\quad
\label{eq:IHM}
\end{eqnarray}
where $\bm{ r}:=i\hat{x}+j\hat{y}$ spans the square lattice, 
$c^\dagger_{\bm{ r},\sigma}$ ($c^{\phantom{\dagger}}_{\bm{ r},\sigma}$)
is the fermionic creation (annihilation)  operator
at site $\bm{ r}$ with spin $\sigma$, and 
$n^{\phantom{\dagger}}_{\bm{ r},\sigma}:=c^\dagger_{\bm{ r},\sigma}c^{\phantom{\dagger}}_{\bm{ r},\sigma}$ is the occupation operator. 
The sum over \mbox{$\langle\bm{ r}\bm{ r'}\rangle$} 
restricts the hopping to nearest-neighbor  sites. 
Initially, the IHM was introduced to describe the 
neutral-ionic transition in mixed-stack organic compounds \cite{Nagaosa1986a}
and was later used for the description of 
 ferroelectric perovskites \cite{Egami1993}.

In 1D, it is well understood that the BI at 
small Hubbard interaction $U$ is separated from quasi-long-range ordered MI at 
large $U$ by an intermediate phase with alternating bond order (BO) 
\cite{Fabrizio1999}. The position of the two transition points 
($U_{c1}$ from BI to BO and $U_{c2}$ from BO to MI)
\cite{Manmana2004,Tincani2009} and the excitation spectrum 
\cite{Hafez2014,Hafez2015} of the model are determined 
quantitatively. We highlight that the transition to the BO
phase is signaled by the softening of an exciton 
\cite{Fabrizio1999,Manmana2004,Tincani2009} 
located at momentum $\pi$ (setting
the lattice constant to unity) \cite{Hafez2014,Hafez2015}.

In 2D, the limiting BI and MI phases are expected 
at low and high $U$, respectively. 
The MI at large $U$ shows long-range antiferromagnetic
order of N\'eel type (AF) \cite{Manousakis1991}.
But existence and properties of  intermediate phases is highly controversial.
A determinant quantum Monte Carlo (DQMC) analysis of small clusters 
identifies an intermediate metallic region without magnetic or bond order 
\cite{Paris2007,*Bouadim2007}. 
Cluster dynamic mean-field theory (DMFT)
indicates an intermediate phase with an incommensurate bond order 
and a finite staggered magnetization \cite{Kancharla2007}.
Single-site DMFT of the half-filled IHM on the Bethe lattice suggests a single
transition point separating the BI from the magnetically ordered MI. 
\cite{Garg2014}.

\begin{figure}[htb]
  \centering
  \includegraphics[width=0.98\columnwidth,angle=0]{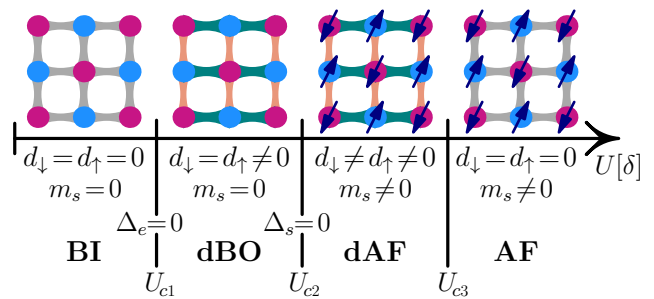}
  \caption{(color online) Scheme of the phase transitions in the 2D IHM. 
The staggered  magnetization is denoted by $m_s$ and the spin-dependent
d-wave bond order parameter by $d_\sigma$. The $S=0$ exciton gap
is $\Delta_e$ and $\Delta_s$ is the spin gap (or $S=1$ exciton) \cite{Hafez2014}.}
  \label{fig:pd}
\end{figure}

We study the 2D IHM by continuous unitary transformations (CUTs) 
\cite{Wegner1994}
realized in real space up to higher orders in the hopping $t$. 
The flow equations are closed using the 
directly evaluated enhanced perturbative CUT (deepCUT) \cite{Krull2012}.
In this way, an effective model is derived in terms of the elementary
fermionic QPs. The virtual processes
are eliminated as in the derivation of a $t$-$J$ model
from a Hubbard model \cite{Harris1967,Chao1977,Hamerla2010,jedra11}. 
The effective model allows us to address the interaction of two of these QPs. 
A particle and a hole attract each other and may form a bound state, an exciton.
The softening of such excitons indicate the instability of the phase
towards a condensation of these excitons.
In this fashion, we determined the BI-to-BO  transition value $U_{c1}$ 
of the 1D IHM within 0.3\% \cite{Hafez2014}. We stress
that already the second order calculation in the hopping $t$ 
yields qualitatively the correct result, i.e., the correct symmetries for
the condensate phase, in spite of the very small energy scales $t^2/U$ 
(relative to $\delta$) driving the transition \cite{Hafez2014,Hafez2015}. 
Thus we adopt the same strategy in 2D
where the CUT has been successfully used before 
\cite{Uhrig2004,Powalski2015}, in particular for binding phenomena 
\cite{Knetter2004}.

We are using the double 1-particle gap $\Delta_c:=2(E_0^{N+1}-E_0^{N})$,
the singlet exciton gap $\Delta_e:=E^N_{1,S=0}-E_0^N$, 
and the spin (triplet) exciton gap 
$\Delta_s:=E^N_{1,S=1}-E_0^N$ where $E_0^N$ stands for the ground state energy 
and $E_{1,S}^N$ for the first excited state 
with total spin $S$ for $N$ electrons. At half-filling, the number of electron  
$N$ equals the number of lattice sites.
In the absence of any electron-electron bound state, $\Delta_c$, being twice the minimum of 
the fermion dispersion, equals the charge gap. We use the term charge gap
because it has been used in previous papers on the IHM 
\cite{Manmana2004,Tincani2009,Hafez2014,Hafez2015}.

Fig.\ \ref{fig:pd} summarizes our key results: Upon increasing $U$ in the BI, the  exciton gap $\Delta_e$ reduces till it vanishes at $U_{c1}$ 
where d-wave bond order (dBO) sets in. In this phase, the $90^\circ$ rotation 
symmetry of the square lattice is broken and the d-wave bond order 
parameter $d_\sigma$ becomes finite. Then, the spin gap $\Delta_s$ closes at 
$U_{c2}$ beyond which the staggered N\'eel magnetization $m_s$ becomes finite
and the  d-wave order parameter $d_\sigma$ starts depending on spin (dAF). 
Eventually, the rotational symmetry is restored at $U_{c3}$ ($d_\sigma=0$)
while the N\'eel order persists (AF).

Technically, we proceed as in 1D \cite{Hafez2014}
by transforming the half-filled IHM into a translationally invariant
problem. On the odd sublattice ($i+j$ odd), we apply the electron-hole transformation 
$T^{(eh)}\!:~c^\dagger_{\bm{r},\sigma}
\rightarrow h^{\phantom{\dagger}}_{\bm{r},\sigma}$. For consistency, we then use for particle and hole creation the same operator 
$f^{(\dagger)}_{\bm{r},\sigma}$. The local phase transformation 
$T^{(l)}\!:~f^{\dagger}_{\bm{r},\sigma} 
\rightarrow e^{i\frac{\pi}{2}(x+y)}e^{-i\frac{\pi}{4}}f^{\dagger}_{\bm{r},\sigma}$
yields
\bearr
H &=& (U-2\delta)/4 \sum_{\bm{ r}} \mathds{1} 
+ t \sum_{<\bm{ r}\bm{ r'}\!>,\sigma} ( f^\dagger_{\bm{ r},\sigma} 
f^\dagger_{\bm{ r}',\sigma} + {\rm h.c.} )\nn 
\\
&+& (\delta - U)/2 \sum_{\bm{ r},\sigma} f^\dagger_{\bm{ r},\sigma} f_{\bm{ r},\sigma}^{\phantom{\dagger}}
+U \sum_{\bm{ r}} f^\dagger_{\bm{ r},\uparrow}f^\dagger_{\bm{ r},\downarrow} 
f_{\bm{ r},\downarrow}^{\phantom{\dagger}} f_{\bm{ r},\uparrow}^{\phantom{\dagger}}.
\quad
\label{eq:IHM_frep}
\eearr
The phase transformation $T^{(l)}$ induces a shift
in momentum space by $\bm{\pi}/2:=(\pi/2,\pi/2)$ for single $f$-fermions.
The electron-hole transformation $T^{(eh)}$ modifies the spin and 
the charge on the odd sublattice according to
\be 
(S^x_{\bm{ r}}, S^y_{\bm{ r}}, S^z_{\bm{ r}}, n^{\phantom{+}}_{\bm{ r}, \sigma}) \longrightarrow 
(-S^x_{\bm{ r}}, +S^y_{\bm{ r}}, -S^z_{\bm{ r}}, 1-n^{\phantom{+}}_{\bm{ r}, \sigma}).
\label{eq:spin_trans}
\ee
Hence, the AF phase in the original $c$-fermions
appears as  translationally invariant ferromagnet in the $f$-fermions.
The staggered magnetization reads
$m_s:=\frac{1}{2}\sum\limits_{\sigma} {\rm sgn}(\sigma)
\langle f^\dagger_{\bm{0},\sigma} f_{\bm{0},\sigma}^{\phantom{\dagger}} \rangle$. 
The d-wave order parameter reads
$d_\sigma^{\phantom{\dagger}}:=
\langle f^{\dagger}_{\bm{0},\sigma}f^{\dagger}_{\bm{\hat{x}},\sigma} \rangle - 
\langle f^{\dagger}_{\bm{0},\sigma}f^{\dagger}_{\bm{\hat{y}},\sigma} \rangle$.

\begin{figure}[t]
  \centering
  \includegraphics[width=0.94\columnwidth,angle=-90]{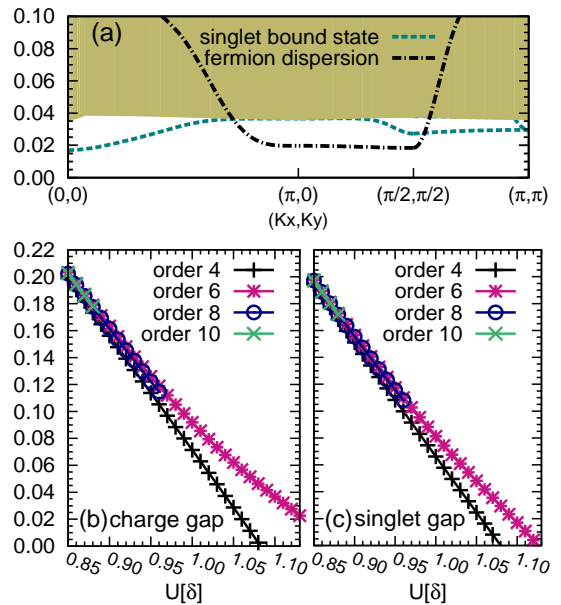}
  \caption{(color online) (a) Low-energy excitations for $t=0.05\delta$ and 
	$U=1.1\delta$ in order 6; charge gap $\Delta_c$ (b) and singlet gap 
	$\Delta_s$ (c) in units of $\delta$ versus $U$.}
  \label{fig:bi}
\end{figure}

We apply the deepCUT to \eqref{eq:IHM_frep} eliminating
processes which let single $f$-fermions and pairs of them 
decay. We disentangle the one-QP and the two-QP subspaces
from the remaining Hilbert space by a suitable change of basis such 
that the 1-QP dispersion
and the dispersion of bound states such as  excitons become accessible.
Technically, we employ the reduced generator \mbox{$n:2$} \cite{Fischer2010}
targeting the 0-, 1-, and 2-QP sector \cite{Krull2012}. 
The sign of the terms in the generator, i.e., the sense of rotation, is chosen 
first according to the change in the number of double occupancy (increase
of double occupancy leads to the same sign in the generator as
in the Hamiltonian). If this number is not changed, the change in the number of 
fermions determines the sign. In this way, we map
the Hamiltonian \reqn{eq:IHM_frep} to the effective Hamiltonian
\be
H_{\rm eff}={\rm E}_0 +
\sum_{\bm{r},\sigma} \sum_{\bm{d}}
\prescript{\sigma \!}{\sigma \!}{\big{[}}{ \mathcal{C}_{1}^{1}\big{]}}^{\!\bm{d}}
 (f^\dagger_{\bm{r},\sigma} f_{\bm{r}+\bm{d},\sigma}^{\phantom{\dagger}} + 
{\rm h.c.}) +H_2^2
\label{eq:IHM_eff}
\ee
where $H_2^2$ stands for interaction terms of two creation and two annihilation operators with coefficients $\mathcal{C}_{2}^{2}$.
The effective coefficients $\mathcal{C}_{1}^{1}$ and $\mathcal{C}_{2}^{2}$ 
are determined from the flow equation $\partial_\ell H=[\eta,H]$
\cite{Wegner1994,Knetter2000,Kehrein2006,Fischer2010,Krull2012}. 
Higher particle interactions are ignored.
The range of processes in \reqn{eq:IHM_eff} is limited by the order in $t$ of the deepCUT. The calculations are restricted to order 6 because
for higher orders overlapping continua prevent a well-defined convergence
of the flow \cite{Fischer2010}. Thus the effective Hamiltonian is not 
fully quantitative, but it is qualitatively correct. For an assessment 
of the accuracy see  Refs.\ \onlinecite{Hafez2014,Hafez2015};
technical aspects related to simplification rules are given in Appendix \ref{app:sec:sr}.

The effective Hamiltonian \reqn{eq:IHM_eff} allows us 
to determine the ground state energy, the 1-QP dispersion, 
and bound pairs of 2 QPs. Due to the particle-hole
transformation $T^{(eh)}$ excitons with odd distance
between their constituents appear in the effective model as pairs of particles.
The eigenvalues in the 2-QP subspace are found for  fixed 
total momentum $\bm{k}$, total spin $S$, and total $S^z$ component $M$
as in the 1D case \cite{Hafez2014}. 
Fig.\ \ref{fig:bi}(a) depicts the low-energy spectrum in the BI phase 
consisting of the fermionic dispersion and a singlet exciton; 
no triplet exciton is found. Note the rather flat dispersion and
lower continuum edge. 

Very interestingly, the singlet exciton takes its minimum at $\bm{k}=(0,0)$. 
In 1D, its minimum was at $k=\pi$ so that its condensation 
led  to \emph{dimerized} bond order \cite{Fabrizio1999,Hafez2014}. 
The fact that the singlet exciton softens at a different momentum
clearly shows that the condensed phase beyond the BI will no longer
be a dimerized BO.  Figs.\ \ref{fig:bi}(b) 
and (c) display the charge and the singlet gap vs.\ $U$. 
Generally, the softening of a bosonic
excitation implies the divergence of the susceptibility 
with the same symmetries $\chi(\omega)$ at $\omega=0$ because 
$\chi(\omega)$ corresponds to the bosonic propagator up to matrix elements.
Thus it has a pole at the boson eigen energy moving to $\omega=0$
upon softening. The direct access to the
energies of bound states avoids the necessity to guess the correct
symmetry of the diverging susceptibility.
In order $4$, the singlet gap $\Delta_e$ vanishes at $U_{c1}=1.08\delta$ 
where the charge gap $\Delta_c$ is still finite at $0.003\delta$. In order $6$, 
we find $U_{c1}=1.126\delta$ with $\Delta_c=0.023\delta$. 
Considering all available orders and comparing to the 1D results \cite{Hafez2014}, 
we estimate that the
transition point $U_{c1}$ lies between $U=1.08\delta$ and $U=1.126\delta$.
The value of the charge gap at the transition point is also expected to be 
between the results in order 4 and order 6, similar to 
1D findings \cite{Hafez2015}.
Increasing $t$ to $0.1\delta$ the transition point
shifts to larger values of $U$, see Appendix \ref{app:sec:Larger_t}.
 
\begin{figure}[t]
  \centering
  \includegraphics[width=0.9\columnwidth,angle=0]{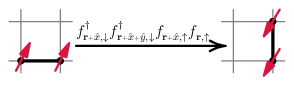}
  \caption{Correlated hopping
$f_{\bm{r}+\hat{x},\bar{\sigma}}^{\dagger} 
f_{\bm{r}+\hat{x} + \hat{y},\bar{\sigma}}^{\dagger}
f_{\bm{r}+\hat{x},\sigma}^{\protect\phantom{\dagger}}
f_{\bm{r},\sigma}^{\protect\phantom{\dagger}}$.
Recall that the spin configuration shown is
a particle-hole singlet in the original $c$-fermions.}
  \label{fig:corr_hopp}
\end{figure}

The key observation is that
the $S=0$ exciton softens \emph{before} the system would become metallic
in analogy to the 1D case, but with \emph{other} symmetries.
The exciton condensation leads to a phase without magnetic order because
the condensing exciton has no spin. The condensation 
does not break translational
invariance because the exciton condenses at zero total moment.
The analysis of the point group properties of the
exciton reveals that its wave function
becomes negative upon spatial rotation by $90^\circ$ so that 
the condensed phase should display the same d-wave symmetry.

The mechanism for electron and hole to form a bound state 
with d-wave symmetry relies on the interaction
$ 
f_{\bm{r}+\hat{x},\bar{\sigma}}^{\dagger} 
f_{\bm{r}+\hat{x} + \hat{y},\bar{\sigma}}^{\dagger}
f_{\bm{r}+\hat{x},\sigma}^{\phantom{\dagger}}
f_{\bm{r},\sigma}^{\phantom{\dagger}}
$ 
representing correlated hopping, see Fig. \ref{fig:corr_hopp}. 
Such terms, occurring already in the second order in $t$ 
with positive sign, have a singlet on a bond in 
$\hat{x}$-direction hop to a bond in $\hat{y}$-direction 
\footnote{Due to the electron-hole transformation, 
the 2-QP singlet state is given by 
$\ket{eh}^{S=0}=\left( \ket{\uparrow\uparrow} + \ket{\downarrow\downarrow} \right)/\sqrt{2}$.}. 
This correlated hopping corresponds
to the well-known three-site terms in the $t$-$J$ model
\cite{Harris1967,Chao1977,Hamerla2010,jedra11}.
If the wave function has the opposite sign on $\hat{y}$-bonds
of those on $\hat{x}$-bonds this correlated hopping
leads to attraction and eventually to a bound state 
with d-wave symmetry.

\begin{figure}[htb]
  \centering
  \includegraphics[width=1.1\columnwidth,angle=-90]{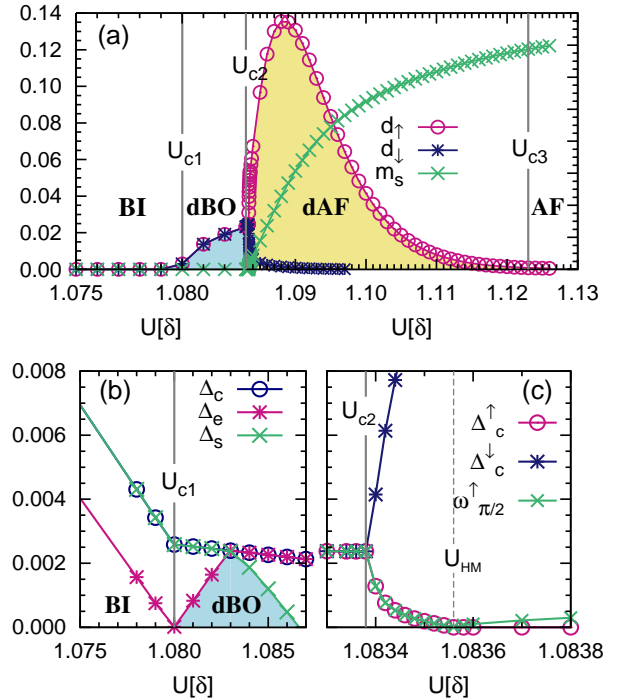}
  \caption{(color online) Staggered magnetization density $m_s$ and 
  d-wave bond order parameter $d_\sigma$ (panel (a)) and 
  charge gap $\Delta_c^\sigma$, singlet gap $\Delta_e$, and spin gap $\Delta_s$ 
  (panels (b) and (c)) in units of $\delta$ versus $U$ for $t=0.05\delta$ at order $4$.}
  \label{fig:cut_mf}
\end{figure}

Similar to the 1D case \cite{Hafez2014}, we describe this condensed phase
by a self-consistent BCS-type
MF theory applied to \eqref{eq:IHM_eff}, see Fig.\ 
\ref{fig:cut_mf}(a) to the left of $U_{c2}$. 
This approach is exact for $U\leq U_{c1}$ where the quantum fluctuations 
are already considered by the CUT. In particular, the 
onset of order occurs precisely where the exciton energy vanishes
as required by consistency. For $U>U_{c1}$ the
MF approach is a good approximation systematically controlled
by the distance $U>U_{c1}$. The order of the calculations is $4$ in 
powers of $t$ and $t=0.05\delta$.
Its bond order $d_\sigma$ consists in the difference of the bonds 
in $x$ and in $y$ direction implying d-wave character. 
We determine the expectation values 
$\langle f^\dagger_{\bm{ 0},\sigma} 
f_{\bm{ d},\sigma}^{\phantom{\dagger}} \rangle$ and 
$\langle f^\dagger_{\bm{ 0},\sigma} f_{\bm{ d},\sigma}^{\dagger} \rangle $
self-consistently, see Appendix \ref{app:sec:mf}, comprising also
the magnetization $m_s$ and the d-wave order parameter $d_{\sigma}$.

The singlet and the triplet excitons \emph{within} the dBO are determined 
by analysing the 2-QP problem neglecting the off-diagonal terms 
linking 2-QP states to states of higher particle number, see Appendix 
\ref{app:sec:mf} for the details. This leads to the gaps displayed in Fig.\ \ref{fig:cut_mf}(b).  
The  exciton gap $\Delta_e$
quickly increases till the exciton ceases to exist because
it merges with the 2-QP continuum.
In parallel, a triplet exciton is formed and softens. 
The results of order $6$, given in Appendix \ref{app:sec:o6}, are qualitatively similar.
Hence the system becomes unstable towards the 
condensation of a magnetic bosons, i.e., towards magnetic ordering.
The softening occurs at momentum $\bm{\pi}$
in the original model so that this triplon condensation is the 
expected N\'eel ordering.

We study the N\'eel ordering by allowing for spin dependent MF solutions
of \eqref{eq:IHM_eff}, i.e., $m_s$ becomes
finite and $d_\uparrow\neq d_\downarrow$. In addition,
the charge gap becomes also spin dependent 
$\Delta^\uparrow_c\neq\Delta^\downarrow_c$.
We find that the d-wave order does not vanish, 
but that it persists so that antiferromagnetism
and d-wave order coexist and a d-wave antiferromagnet (dAF)
forms. This solution occurs at and beyond $U_{c2}\approx 1.083\delta$ 
as shown in Fig.\ \ref{fig:cut_mf}(a) and in the zoom (c). 
\footnote{The value of $U_{c2}$ is slightly
lower than the value where $\Delta_s$ vanishes in Fig.\
\ref{fig:cut_mf}(b) because the off-diagonal terms 
neglected in the calculation of $\Delta_s$ contribute 
in the MF solution favoring N\'eel ordering.}

So far, we constructed the phase diagram following the
spontaneous symmetry breaking indicated by soft excitons. In the dAF phase,
the symmetry is  strongly reduced. For large $U$, however, we 
expect N\'eel ordering without d-wave components because the
Hubbard model maps to the Heisenberg model since the ionic alternation
is only a small perturbation for $U\gg\delta$. The 
MF solution retrieves this feature. The
BO parameters $d_\sigma$ quickly decrease on increasing $U$.
For $U > 1.123\delta$, $d_\uparrow$ is smaller than $10^{-3}$.
We conclude that there is a $U_{c3}$ beyond which the spatial rotation
is restored as a symmetry of the IHM: the dAF phase becomes 
a N\'eel ordered AF phase. The precise value of $U_{c3}$
is difficult to pinpoint because $d_\uparrow(U)$ is
consistent with an exponential decrease $\propto \exp(C/(U-U_{c3}))$ 
with $U_{c3}\approx 1.24\delta$. This resembles a Kosterlitz-Thouless type of 
transition as also proposed in Ref.\ \onlinecite{Kancharla2007}. But 
the numerics is too involved and the energy scales too small to identify 
the decrease of $d_\uparrow(U)$ unambiguously.

In Fig.\ \ref{fig:cut_mf}(c), we also plot the spin-dependent 
charge gaps in the vicinity of $U_{c2}$. Beyond $U_{c2}$, 
$\Delta_c^\downarrow$ increases quickly, but $\Delta_c^\uparrow$ decreases 
and hits zero at a half-metallic point $U_{\rm HM}$. 
Such half-metallic points occurred in previous analyses 
already \cite{Garg2014}. The spin-up charge gap is located at momentum 
$\bm{k}=\bm{\pi}/2$ up to $U_{\rm HM}$, but for $U>U_{\rm HM}$ 
it moves to incommensurate momenta at the line $k_x=k_y$ and remains zero. 
But we cannot exclude that the zero $\Delta_c^\uparrow$ for $U>U_{\rm HM}$ is 
an artefact of our approach. Any small deviation from the
particle-conserving effective Hamiltonian  \reqn{eq:IHM_eff}
would spoil this feature. Thus we also included $\omega^{\uparrow}_{\bm{\pi}/2}$
in Fig.\ \ref{fig:cut_mf}(c) displaying a more generic behavior.

Summarizing, we studied unconventional phases in the ionic
Hubbard model in 2D between the known band insulator at weak interaction 
and the N\'eel ordered Mott insulator at strong interaction.
We found bond order which spontaneously breaks
the orientational symmetry: d-wave bond order (dBO). Even a region of 
coexistence of bond order and N\'eel order is identified: d-wave bond order and 
antiferromagnetism (dAF). Thereby, we provide clarification for the so far ambiguous evidence concerning the influence of strong interactions on
band insulators. We stress that no weak-coupling instabilities indicate 
the existence of these  phases. This underlines that these phases 
are inherent to strong coupling \cite{Schulz1989}.
Against this background, we think
that the single-site DMFT \cite{Garg2014} does not capture the 
spatial correlations indispensable for d-wave ordering and the DQMC 
\cite{Paris2007,*Bouadim2007} considers too small clusters at too 
large temperatures to capture the small energy scales driving the transitions found. 
The cluster DMFT \cite{Kancharla2007} also advocated bond order, 
but combined with dimerization and incommensurability. Since the rotational
symmetry was not allowed to be broken no d-wave ordering could
be detected.

We emphasize that the scenario discovered here is different
from the bond ordered phases with d-wave symmetry discussed in the literature
so far. These previous scenarios imply either a kind of alternation
\cite{Schulz1989,Vojta2008} or incommensurability \cite{Metlitski2010a,Sachdev2013} 
and they usually occur \emph{away} from half-filling in contrast to our finding. 
The dBO phase presently advocated induces no dimerization and breaks only 
the rotational symmetry of the square lattice.

The conclusions are based on effective models 
expressed in elementary excitations. This approach was 
successful in 1D. No {\it ad hoc} assumptions about 
broken symmetries are made, but instabilities are systematically
deduced from softening bound states, namely singlet
and triplet excitons, the latter being the spin excitations.
The quantum numbers and symmetries of the excitonic wave function
implies the symmetry of the phase to which the instable phase evolves. 
In this way, two continuous phase transitions at $U_{c1}$ and
$U_{c2}$ are identified. Based on symmetry considerations, we expect the 
first transition to be in the Ising and the second transition to be in the 
$O(3)$  universality class.

We consider the obtained scenario to be complete because 
the d-wave order quickly decreases in the dAF phase so that
only N\'eel order remains
beyond a third transition interaction $U_{c3}$ as expected 
for large interaction $U$.

The findings indicate for which types of correlations
one has to look in order to identify intermediate
phases in the wide research field between
band insulating and Mott insulating phases, for instance 
on various lattices.

\begin{acknowledgments}
We are grateful to the Helmholtz Virtual Institute 
``New states of matter and their excitations'' for financial
support.
\end{acknowledgments}

\appendix

\section{Simplification Rules}
\label{app:sec:sr}

Here we present the simplification rules used to 
realize the comprehensive CUTs, cf.\ Ref.\ \onlinecite{Krull2012}.

The basic {\it a-posteriori} and {\it a-priori} simplification rules (SRs) introduced in 
Ref. \onlinecite{Hafez2014} can be used for the IHM on the square lattice because they are lattice independent. 
In addition, 
we have implemented an extended {\it a-posteriori} SR.
The aim is to estimate whether a monomial $A$ contributes to the targeted quantities 
up to a specific order, for example $n$, or not.
We explain the extended {\it a-posteriori} SR for the creation operators in $A$. 
The annihilation part can be analyzed in the same way. 

Similar to the 1D case, one can treat spin-up and spin-down operators separately \cite{Hafez2014}.
From the position of creation operators with spin $\sigma$, we form a cluster (graph). 
We consider a vertex for each lattice site and add an edge between two vertices if the 
corresponding sites are adjacent. The degree of a vertex is defined as the number of edges 
linked to the vertex.
The cluster is divided into a set of linked subclusters. Let us consider one of the subclusters 
and denote it by $\mathcal{C}$.
In the first step, we need to underestimate the number of commutations $K[\mathcal{C}]$ necessary 
to cancel the arbitrary linked cluster $\mathcal{C}$. 

The size of $\mathcal{C}$ is reduced 
in an {\it iterative} way. We eliminate each vertex with degree 1 and the vertex it is linked to. 
This costs one commutation for each of these pairs to be omitted. Let us suppose that after 
$p^{\phantom{\mathcal{C}}}_{\mathcal{C}}$
commutations no vertex with degree 1 is left.
This reduces the linked cluster $\mathcal{C}$ to isolated vertices and/or 
to the linked subclusters $\mathcal{C}'$ which has no vertex with degree 1. 
Each isolated vertex requires one commutation to be canceled. We denote by 
$s^{\phantom{\mathcal{C}}}_{\mathcal{C}}$ the number of isolated vertices.
One also needs  
$ \left\lceil \dfrac{v_{\mathcal{C}'}^{\phantom{r}}}{2} \right\rceil $ commutations to cancel the 
subcluster $\mathcal{C}'$ which has $v_{\mathcal{C}'}^{\phantom{r}}$ vertices. 
$\left\lceil x \right\rceil$ is defined as smallest integer not less than $x$.
Hence, we obtain
\bseq
\label{eq:sr:partial_comm}
\begin{align}
\label{eq:sr:partial_comm:a}
K[\mathcal{C}] \geq \widetilde{K}\left[ \mathcal{C} \right]
:=& \  p^{\phantom{\mathcal{C}}}_{\mathcal{C}}+s^{\phantom{\mathcal{C}}}_{\mathcal{C}}
+\sum_{\mathcal{C}'} \left\lceil \frac{v_{\mathcal{C}'}^{\phantom{\mathcal{C}}}}{2} \right\rceil 
\\
\label{eq:sr:partial_comm:b}
\geq & \  p^{\phantom{\mathcal{C}}}_{\mathcal{C}}+s^{\phantom{\mathcal{C}}}_{\mathcal{C}}
+ \left\lceil \frac{v_{\mathcal{C}}^{r}}{2} \right\rceil,
\end{align}
\eseq
where $v^r_{\mathcal{C}}:=\sum\limits_{\mathcal{C}'} v_{\mathcal{C}'}^{\phantom{\mathcal{C}'}}$.
One can use the simpler Eq. \reqn{eq:sr:partial_comm:b} instead of \reqn{eq:sr:partial_comm:a} 
for $\widetilde{K}\left[ \mathcal{C} \right]$.
There should be at least two subclusters $\mathcal{C}'$ each one with an odd number of vertices 
to change the ``equal'' sign in \reqn{eq:sr:partial_comm:b} into ``greater than''. For the square lattice, 
this means that the initial linked cluster $\mathcal{C}F$ should contain at least 16 (=7+7+2) vertices.

\begin{figure}[ht]
  \centering
  \includegraphics[width=0.9\columnwidth,angle=0]{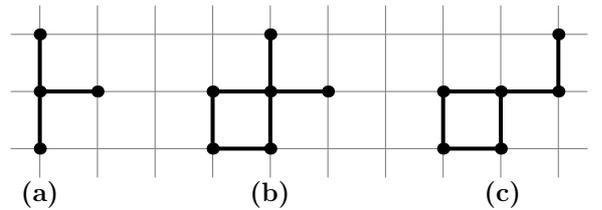}
  \caption{Three kinds of linked clusters on the square lattice.}
  \label{fig:cluster}
\end{figure}

One can check that 
Eq. \reqn{eq:sr:partial_comm} simplifies to the 1D counterpart where no subcluster $\mathcal{C}'$ 
can exist \cite{Hafez2014}. 
As example, let us examine the three clusters in Fig. \ref{fig:cluster}. 
We find 
$p^{\phantom{\mathcal{C}}}_{\mathcal{C}}=1$, $s^{\phantom{\mathcal{C}}}_{\mathcal{C}}=2$, 
and no $\mathcal{C}'$ for Fig. \ref{fig:cluster}(a),  
$p^{\phantom{\mathcal{C}}}_{\mathcal{C}}=2$, $s^{\phantom{\mathcal{C}}}_{\mathcal{C}}=2$, 
and no $\mathcal{C}'$ for Fig. \ref{fig:cluster}(b), 
and finally 
$p^{\phantom{\mathcal{C}}}_{\mathcal{C}}=1$, $s^{\phantom{\mathcal{C}}}_{\mathcal{C}}=0$, 
and one subcluster $\mathcal{C}'$ with $v_{\mathcal{C}'}=4$ for Fig. \ref{fig:cluster}(c). 

The total number of commutations $\widetilde{K}^c_0$ 
required to cancel all the creation operators of $A$ is given by
\be
\widetilde{K}^c_0:=\sum_{\mathcal{C}} \widetilde{K}\left[ \mathcal{C} \right]
\label{eq:sr:total_comm}
\ee
where the sum runs over all linked spin-up and spin-down clusters of monomial $A$.
Similar to the rules presented in Ref. \onlinecite{Krull2012}, 
Eq. \reqn{eq:sr:total_comm} can be extended to $\widetilde{K}^c_q$ where sectors up to 
$q$ quasiparticles are targeted. The aim is to keep $q$ operators such that the maximal 
order of $A$ is overestimated \cite{Krull2012}. We first keep the operators which can save 
one commutation. The total number of these operators is given by 
\be
d:=\sum \limits_{\mathcal{C}} d_{\mathcal{C}}^{\phantom{\mathcal{C}}}
:=\sum \limits_{\mathcal{C}} 
\left( s_{\mathcal{C}}^{\phantom{\mathcal{C}}} + v_{\mathcal{C}}^r \ {\rm mod}\ 2 \right),
\ee
where we suppose that \reqn{eq:sr:partial_comm:b} holds for $\widetilde{K}[\mathcal{C}]$. 
The remaining operators (if $q>d$) can save one commutation per pair. Therefore, 
we find
\be
\widetilde{K}^c_q={\rm max}\left( \widetilde{K}^c_0 - \left\lfloor \frac{q+d'}{2} \right\rfloor, 0 \right),
\ee
where $d':={\rm min}(d,q)$. In the same way, one can analyze the annihilation part of 
the monomial $A$ to find $\widetilde{K}^a_q$. The monomial $A$ can be discarded if 
\be
O_{\rm min}(A) > n - \widetilde{K}^c_q - \widetilde{K}^a_q,
\ee
where $O_{\rm min}(A)$ is the minimal order of $A$ \cite{Krull2012}.

\section{Results for Larger Hopping}

\begin{figure}[t]
  \centering
  \includegraphics[width=1.0\columnwidth,angle=-90]{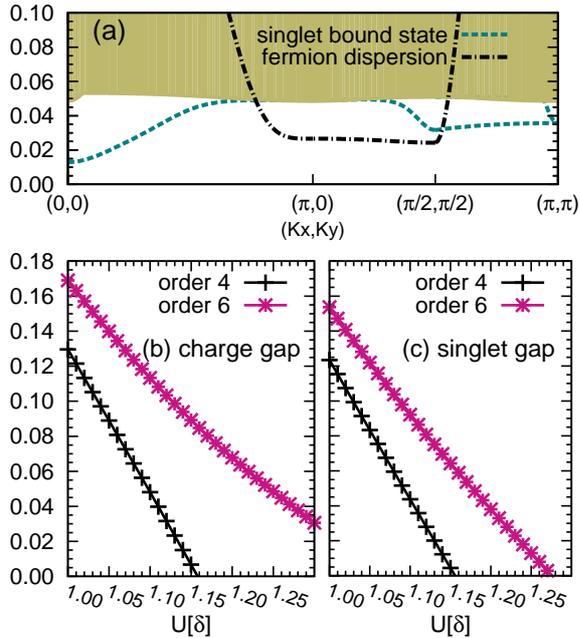}
  \caption{(a) The same as Fig. \ref{fig:bi}(a) in the main text, but for $t=0.1\delta$ and $U=1.25\delta$. 
(b) and (c) are also the same as Fig. \ref{fig:bi}(b) and (c) in the main text, but for $t=0.1\delta$.}
  \label{fig:t0.1}
\end{figure}

In order to show that the results obtained in the main text
also apply for larger values of the hopping.
Fig.\ \ref{fig:t0.1}(a) represents the same quantities 
as Fig. \ref{fig:bi}(a) but for $t=0.1\delta$ and $U=1.25\delta$. 
Figs.\ \ref{fig:t0.1}(b) and (c) are  the same as Figs. \ref{fig:bi}(b) and (c), but for $t=0.1\delta$. 
One can see that by increasing $t$ from $t=0.05\delta$ to $t=0.1\delta$, the position of the first transition point is 
shifted to the larger values of $U$. The value of the charge gap at the transition point is also increased compared 
to the case for $t=0.05\delta$. 
This behaviour is similar to what is found for the 1D IHM by DMRG \cite{Manmana2004}. Larger hopping increases the energy scale on which
the relevant physics takes place in the ionic Hubbard model (IHM).

\label{app:sec:Larger_t}

\section{BCS-Type Mean-Field Analysis}
\label{app:sec:mf}

The Hamiltonian \reqn{eq:IHM_eff} is normal-ordered with respect 
to the bilinear part using Wick's theorem. 
The bilinear Hamiltonian is diagonalized in momentum space by a Bogoliubov transformation. 
The relevant expectation values are given by
\bseq
\label{eq:mf}
\begin{align}
 \langle f^\dagger_{\bm{ 0},\sigma} f_{\bm{ d},\sigma}^{\phantom{\dagger}} \rangle &= 
\frac{\delta_{\bm{d},\bm{0}}}{2}-\frac{1}{2\pi^2}\int_{(\scalebox{0.75}[1.0]{-}\frac{\pi}{2},0)}^{(\frac{\pi}{2},\pi)} \!
\frac{{\widetilde{c}}^{\sigma}_{\bm{k}}}{\omega^{\sigma}_{\bm{k}}} \cos(\bm{k} \! \cdot \! \bm{d}) \, {\rm d}\bm{k} , \\
 \langle f^\dagger_{\bm{ 0},\sigma} f_{\bm{ d},\sigma}^{\dagger} \rangle &=
-\frac{1}{2\pi^2}\int_{(\scalebox{0.75}[1.0]{-}\frac{\pi}{2},0)}^{(\frac{\pi}{2},\pi)} \!
\frac{\widetilde{\Gamma}^{\sigma}_{\bm{k}}}{\omega^{\sigma}_{\bm{k}}} \sin(\bm{k} \! \cdot \! \bm{d}) \, {\rm d}\bm{k},
\end{align}
\eseq
where 
$\widetilde{c}^{\sigma}_{\bm{k}}:=c^{\sigma}_{\bm{0}}+2\sum\limits_{\bm{d}} 
c^{\sigma}_{\bm{d}}\cos(\bm{k} \! \cdot \! \bm{d})$, 
$\widetilde{\Gamma}^{\sigma}_{\bm{k}}:=
2\sum\limits_{\bm{d}} \Gamma^{\sigma}_{\bm{d}}\sin(\bm{k} \! \cdot \! \bm{d})$, and 
$\omega^{\sigma}_{\bm{k}}:=
\sqrt{ {\widetilde{c}}^{\sigma}_{\bm{k}}{}^2+
{\widetilde{\Gamma}}^{\sigma}_{\bm{k}}{}^2}$.

The coefficients  $c^{\sigma}_{\bm{d}}$ and $\Gamma^{\sigma}_{\bm{d}}$ are defined as prefactors of normal-ordered hopping and Bogoliubov 
operators over the distance $\bm{d}$, respectively. Conservation of total charge guarantees that no hopping term over odd and 
no Bogoliubov term over even distances occurs.
One can omit the spin indices from Eqs.\ \reqn{eq:mf} because the SU(2)
symmetry is preserved in the d-wave bond order (dBO) phase. 
The general spin-dependent form helps us to analyze also phases 
with broken spin symmetry.
After some standard calculations, the Hamiltonian \reqn{eq:IHM_eff} be rewritten  as 
 \begin{widetext}
\be
H_{\rm eff}=\widetilde{\rm E}_0 \mathds{1}+
\sum_{\bm{k},\sigma} \omega_{\bm{k}} \lambda^\dagger_{\bm{k},\sigma} 
\lambda_{\bm{k},\sigma}^{\phantom{\dagger}}
+ \sum_{\substack{\bm{k} \bm{q}_i \bm{q}_f}} 
\sum_{\substack{\sigma_1, \sigma_2 
\\ 
\beta_1, \beta_2}}
\prescript{\beta_2 \beta_1 \!}{\sigma_2 \sigma_1 \!}{\big{[}}{ \widetilde{\mathcal{C}}_{2}^{2}}{\big{]} }^{\bm{k} \bm{q}_f}_{\bm{q}_i}
 \lambda^\dagger_{\bm{k}/2+\bm{q}_f,\beta_2} 
\lambda^\dagger_{\bm{k}/2 - \bm{q}_f,\beta_1} 
\lambda_{\bm{k}/2+\bm{q}_i,\sigma_2}^{\phantom{\dagger}} 
\lambda_{\bm{k}/2-\bm{q}_i,\sigma_1}^{\phantom{\dagger}} + \cdots,
\label{eq:sr:IHM_mom}
\ee
 \end{widetext}
where the operator $\lambda^{(\dagger)}_{\bm{k},\sigma}$ is defined such that it  diagonalizes the non-interacting part of the Hamiltonian. 
The coefficients of the  interaction potential in momentum space
$\widetilde{\mathcal{C}}_{2}^{2}$ 
are given in terms  of the real space interactions $\mathcal{C}_{2}^{2}$ and 
the hopping coefficients $\widetilde{c}^{\phantom{\dagger}}_{\bm{k}}$ 
and the Bogoliubov coefficients 
$\widetilde{\Gamma}^{\phantom{\dagger}}_{\bm{k}}$.
The ``$\cdots$'' stand for the quartic off-diagonal interactions 
$\widetilde{H}^3_{1} \propto \lambda^{\dagger} \lambda^{\dagger} 
\lambda^{\dagger} \lambda^{\phantom{\dagger}}$ 
and 
$\widetilde{H}^4_{0} \propto \lambda^{\dagger} 
\lambda^{\dagger} \lambda^{\dagger} \lambda^{\dagger}$
and their hermitian conjugates. 

Taking the ground-state energy of the mean-field approximation
as the true ground-state energy neglects the terms
$\widetilde{H}^4_{0} + \widetilde{H}^0_{4}$. They are systematically
controlled by $U-U_{c1}$, i.e., they are indeed small in the
parameter regime considered.

Considering  the 1-particle dispersion $\omega_{\bm{k}}$ in 
Eq.\ \reqn{eq:sr:IHM_mom} and/or the interactions in the 2-particle 
sector above the mean-field solution, we also neglect the terms 
$\widetilde{H}^3_{1}+ \widetilde{H}^1_{3}$ which
are again systematically controlled by $U-U_{c1}$. 
The singlet and the triplet 2-particle bound states in Fig. \ref{fig:cut_mf}(b) of the main 
text are determined by  exact diagonalization in the 2-particle subspace for fixed total momentum  $\bm{k}$, total spin $S$, and total magnetic number $M$. 
The remaining quantum number is the relative momentum $\bm{q}$ which is not conserved.

There is, however, a subtle point to be noticed. In the 2-particle state 
$\ket{\bm{k};\bm{q}}_{\sigma_1\sigma_2}
:=\lambda^\dagger_{\bm{k}/2 + \bm{q},\sigma_1} 
\lambda^\dagger_{\bm{k}/2 - \bm{q},\sigma_2}\ket{0}$,  
states with two electrons (two holes) and one electron and one hole are mixed. 
In order to stay in the half-filled case, we define
$\ket{\bm{k};\bm{q}}_{\sigma_1\sigma_2}^{Q=0}
:=\frac{1}{\sqrt{2}}(\ket{\bm{k};\bm{q}}_{\sigma_1\sigma_2} 
  +  \ket{\bm{k};\bm{\pi} \scalebox{0.75}{$-$} \bm{q} }_{\sigma_2\sigma_1} )$ which has the net total charge $Q=0$.
The relative momentum $\bm{q}$ is now restricted, e.g., to \mbox{$0<q_x<\pi$} and \mbox{-$\frac{\pi}{2}<q_y<\frac{\pi}{2}$}. 
We employ triplet states ($S=1$) with specific polarizations rather than specific magnetic numbers.  The triplet states are given by 
\mbox{$\ket{t_x}=\frac{1}{\sqrt{2}}(\ket{\uparrow\downarrow}+\ket{\downarrow\uparrow})$}, 
\mbox{$\ket{t_y}=\frac{-i}{\sqrt{2}}(\ket{\uparrow\downarrow}-\ket{\downarrow\uparrow})$}, 
\mbox{$\ket{t_z}=\frac{1}{\sqrt{2}}(\ket{\uparrow\uparrow}-\ket{\downarrow\downarrow})$}, 
and the singlet state by 
\mbox{$\ket{s}=\frac{1}{\sqrt{2}}(\ket{\uparrow\uparrow}+\ket{\downarrow\downarrow})$}.

Constructing the Hamiltonian matrix in momentum space is time consuming compared to the real space diagonalization. 
This is the case because for each $\bm{q}_i$ and $\bm{q}_f$ in \reqn{eq:sr:IHM_mom}, one has to perform three summations over weighted real space interaction coefficients to calculate 
$\prescript{\beta_2 \beta_1 \!}{\sigma_2 \sigma_1 \!}{\big{[}}{ \widetilde{\mathcal{C}}_{2}^{2}}{\big{]} }^{\bm{k} \bm{q}_f}_{\bm{q}_i}$.

\section{Results in Order $t^6$}
\label{app:sec:o6}

\begin{figure}[ht]
  \centering
  \includegraphics[width=1.0\columnwidth,angle=-90]{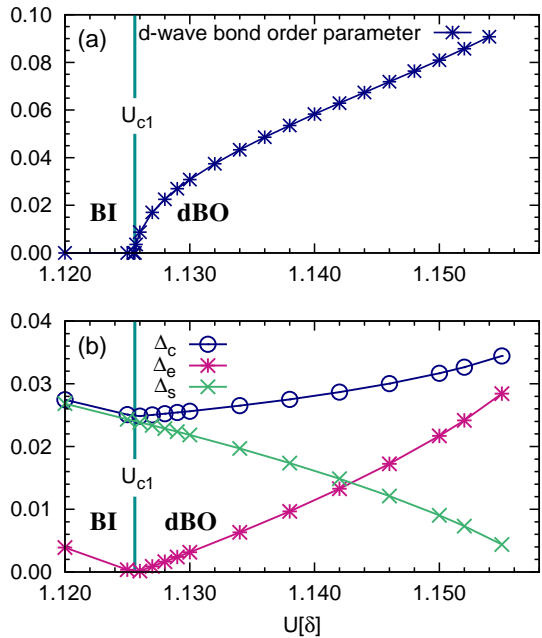}
  \caption{The same as in Fig. \ref{fig:cut_mf}(a) for $U<U_{c2}$ and Fig. \ref{fig:cut_mf}(b), 
	but in order 6 of the hopping.}
  \label{fig:order6}
\end{figure}

In order to underline that the analysis in the main text is supported
by results in higher order of the hopping.
 Figs.\ \ref{fig:order6}(a) and (b) represent the same quantities as Fig. \ref{fig:cut_mf}(a) for $U<U_{c2}$ 
and Fig. \ref{fig:cut_mf}(b), but in order 6. Divergence of the flow equations prevents us to consider interactions beyond $U=1.155\delta$, 
so that we can not access the full phase diagram.

\section*{References}

%

\end{document}